# EFFECT OF THE DIELECTRIC-CONSTANT MISMATCH AND MAGNETIC FIELD ON THE BINDING ENERGY OF HYDROGENIC IMPURITIES IN A SPHERICAL QUANTUM DOT

ARAM KH. MANASELYAN[*], ALBERT A. KIRAKOSYAN

*Department of Solid State Physics, Yerevan State University, Al. Manookian 1, Yerevan 375025, Armenia*

**Abstract**

Within the effective mass approximation and variational method the effect of dielectric constant mismatch between the size-quantized semiconductor sphere, coating and surrounding environment on impurity binding energy in both the absence and presence of a magnetic field is considered. The dependences of the binding energy of a hydrogenic on-center impurity on the sphere and coating radii, alloy concentration, dielectric-constant mismatch, and magnetic field intensity are found for the $GaAs - Ga_{1-x}Al_xAs - AlAs$ (or vacuum) system.

## 1. Introduction

The impurity states in low-dimensional semiconductor heterostructures have been a subject of extensive investigation in basic and applied research [1-3]. The study of hydrogenic impurity-related properties in quantum dots (QD) has been extensively reported in the last few years [4-7]. QD are nowadays considered to be the limit of electronic confinement, and have been widely used in opto-electronic devices and as laser systems. Thus, an understanding of the nature of impurity states in QD is important, since the presence of impurities can dramatically alter the properties and performance of a quantum device.

Impurity states in QD's have been studied in many papers [8-17] where within the effective-mass approximation and variational procedure the binding energies of on- and off-center hydrogenic impurities are calculated for different confining potentials and different shapes of the dots.

The behavior of energy levels of shallow impurities in QD's in the presence of a magnetic field has been investigated in several papers [18-26]. It was shown that in case of an on-center donor, the magnetic field enhances the binding energy and in case of an off-center donor, the effect of the magnetic field on the binding energy is somewhat complicated.

The effect of an electric field on the electronic states and binding energy of a hydrogenic impurity in QD's is considered in [21,27], and the uniaxial stress dependence of the binding energy of shallow donor impurities in a parallelepiped-shaped QD is studied in [28]. It was shown that the binding energy increases with increasing stress of the QD and also with the

---

[*] Corresponding author. Tel.: +3741-553-246

*E-mail address:* amanasel@www.physdep.r.am (A.Kh. Manaselyan).

proximity of the impurity to the center of the QD.

Thus, on the basis of the results obtained in the mentioned works, one can conclude that the impurity states in QD are rather sensitive to changes in sizes, geometrical shape and composition, as well as to the effects of external fields.

Note, that in most of the works devoted to the calculation of the binding energy in QD, the dielectric constant (DC) mismatch between the components of the system, or of the system and surrounding medium, is neglected. The results obtained for QW's [29-31] and QWW's [32,33] testify that a decrease in both the dimensionality and characteristic sizes of the system leads to considerable errors when the dielectric inhomogeneity of the system is neglected.

It is necessary to note that in [34,35] the binding energy calculations are performed taking into account both the spacial variation of dielectric screening and the dielectric mismatch for a spherical QD of GaAs coated by $Ga_{1-x}Al_xAs$. It was shown that the impurity binding energies increase noticeably, especially when the radius of the QD is small.

In this paper we report the calculation of the binding energy of a hydrogenic impurity located at the center of a GaAs sphere with $Ga_{1-x}Al_xAs$ coating in the environment (AlAs or vacuum) surrounding the system. In the presence of a magnetic field the calculations are done using the effective-mass approximation and a variational method within the framework of a staircase infinitely deep (SIW) potential well model with regard of DC mismatches of the sphere, coating and surrounding medium.

## 2. Electronic states in QD in a magnetic field

Within the framework of an effective-mass approximation the Hamiltonian of an electron, in a system consisting of a sphere of radius $R_1$ and coating of radius $R_2$, in the presence of a magnetic field, can be written as

$$\hat{H} = \frac{1}{2\mu(r)}\left(\hat{\vec{p}} + \frac{e}{c}\vec{A}\right)^2 + V(r), \qquad (1)$$

where the effective mass of the electron $\mu(r)$ assumes different values in the sphere and coating:

$$\mu(r) = \begin{cases} \mu_1, & r \leq R_1, \\ \mu_2, & R_1 \leq r \leq R_2, \end{cases} \qquad (2)$$

$\vec{A}$ is the magnetic-field vector potential, and $V(r)$ is the confining potential. Within the framework of the SIW model, $V(r)$ is given by

$$V(r) = \begin{cases} 0, & r \leq R_1, \\ V_0, & R_1 \leq r \leq R_2, \\ \infty, & r > R_2, \end{cases} \qquad (3)$$

where $V_0$ is the value of the potential energy jump at the boundary of the sphere and the coating layer. For a uniform magnetic field $\vec{B}(0,0,B)$, we can write $\vec{A} = [\vec{B}\vec{r}]/2$.

Introducing the Bohr radius $a_B = \chi_1 \hbar^2 / \mu_1 e^2$ as the unit of length, and the effective Ridberg $E_R = \mu_1 e^4 / 2\hbar^2 \chi_1^2$ as the unit of energy, where $\chi_1$ is the dielectric constant of the sphere, the Hamiltonian (1) can be represented as

$$H_i = H_{0i} + H'_i \qquad (i = 1, 2), \tag{4}$$

where $H_{0i}$ and $H'_i$ in spherical coordinates are given by

$$\hat{H}_{01} = -\Delta - i\gamma \frac{\partial}{\partial \varphi}, \qquad H'_1 = \frac{1}{4}\gamma^2 x^2 \sin^2\theta, \tag{5}$$

$$\hat{H}_{02} = -\frac{\mu_1}{\mu_2}\Delta - i\frac{\mu_1}{\mu_2}\gamma \frac{\partial}{\partial \varphi} + v_0, \qquad H'_2 = \frac{\mu_1}{\mu_2}\frac{1}{4}\gamma^2 x^2 \sin^2\theta, \tag{6}$$

where $\gamma = e\hbar B / \mu_1 c E_R$ is a dimensionless measure of the magnetic field, $v_0 = V_0 / E_R$.

In Eqs. (5) and (6) $H'_1$ and $H'_2$ are considered to be the perturbations which are comparable with $H_{01}$ and $H_{02}$, respectively.

The eigenfunctions of the Hamiltonian (1) in the absence of a magnetic field ($\gamma = 0$) are:

$$\psi_{n,l,m}(x,\theta,\varphi) = \frac{C_1}{\sqrt{x}} Y_{l,m}(\theta,\varphi) \begin{cases} J_{l+1/2}(\alpha_{n,l}x), & x \leq x_1, \\ C_2 I_{l+1/2}(\beta_{n,l}x) + C_3 K_{l+1/2}(\beta_{n,l}x), & x_1 \leq x \leq x_2, \end{cases} \tag{7}$$

where $Y_{l,m}$ are the normalized spherical functions, $\hbar l\,(l = 0,1,2...)$ is the momentum, $\hbar m\,(m = 0,\pm 1, \pm 2..., \pm l)$ is the projection of the momentum along the $z$ direction, $C_1$ is the normalization constant of the wave function,

$$\alpha_{n,l,m}^2 = \varepsilon_{n,l,m} - m\gamma, \qquad \beta_{n,l,m}^2 = \frac{\mu_2}{\mu_1}(v_0 - \varepsilon_{n,l,m}) + m\gamma, \tag{8}$$

$$C_2 = \frac{J_{l+1/2}(\alpha_{n,l,m}x_1) K_{l+1/2}(\beta_{n,l,m}x_2)}{I_{l+1/2}(\beta_{n,l,m}x_1) K_{l+1/2}(\beta_{n,l,m}x_2) - K_{l+1/2}(\beta_{n,l,m}x_1) I_{l+1/2}(\beta_{n,l,m}x_2)}, \tag{9}$$

$$C_3 = -\frac{J_{l+1/2}(\alpha_{n,l,m}x_1) I_{l+1/2}(\beta_{n,l,m}x_2)}{I_{l+1/2}(\beta_{n,l,m}x_1) K_{l+1/2}(\beta_{n,l,m}x_2) - K_{l+1/2}(\beta_{n,l,m}x_1) I_{l+1/2}(\beta_{n,l,m}x_2)}, \tag{10}$$

$J_{l+1/2}$, $I_{l+1/2}$ and $K_{l+1/2}$ are the spherical Bessel functions of order $l+1/2$ of the first kind and modified first- and third-kind, respectively [36].

The energy levels are determined from the continuity condition of the logarithmic derivative of the wave function at $r = R_1$. For the ground state ($l = 0, m = 0$, $\alpha_{100} \equiv \alpha; \beta_{100} \equiv \beta$) it has the form:

$$-\frac{\mu_2}{\mu_1}\frac{\alpha J_{3/2}(\alpha x_1)}{J_{1/2}(\alpha x_1)} = \beta \frac{C_2 I_{3/2}(\beta x_1) - C_3 K_{3/2}(\beta x_1)}{C_2 I_{1/2}(\beta x_1) + C_3 K_{1/2}(\beta x_1)}. \tag{11}$$

According to the improved version of perturbation theory [37], which provides a method for treating the problems involving a very large "perturbation" term in the Hamiltonian, the wave function of Eqs. (4) can be derived as

$$\psi^{(1)} = \frac{N_1}{\sqrt{x}} e^{-\frac{1}{4}\chi^2 \sin^2\theta} e^{-\lambda_1 x} \begin{cases} J_{1/2}(\alpha x), & x \leq x_1, \\ C_2 I_{1/2}(\beta x) + C_3 K_{1/2}(\beta x), & x_1 \leq x \leq x_2. \end{cases} \tag{12}$$

In Eq. (12) $\lambda_1$ is the variational parameter, $N_1 = (2\pi A_1)^{-1/2}$ is the normalization constant of the wave function, and $A_1 = D_1 + D_2$, where

$$D_1 = \int_0^{x_1}\int_0^{\pi} x \sin\theta\, J_{1/2}^2(\alpha x) e^{-\frac{1}{2}\chi^2 \sin^2\theta} e^{-2\lambda_1 x} dx d\theta, \tag{13}$$

$$D_2 = \int_{x_1}^{x_2}\int_0^{\pi} x \sin\theta\, [C_2 I_{1/2}(\beta x) + C_3 K_{1/2}(\beta x)]^2 e^{-\frac{1}{2}\chi^2 \sin^2\theta} e^{-2\lambda_1 x} dx d\theta. \tag{14}$$

The ground state energy of an electron in QD is given by

$$E_1 = 2\pi \int_0^{x_1}\int_0^{\pi} \psi^{(1)} \hat{H} \psi^{(1)} x^2 \sin\theta\, dx d\theta + 2\pi \int_{x_1}^{x_2}\int_0^{\pi} \psi^{(1)} \hat{H} \psi^{(1)} x^2 \sin\theta\, dx d\theta. \tag{15}$$

Using Eqs. (5), (6), (9), (10), (12) and (15), after some transformations, we get for the ground state energy:

$$\frac{\varepsilon_1}{E_R} = \alpha^2 + \frac{1}{A_1}\left\{(\lambda_1^2 + \gamma)\left(D_1 + \frac{\mu_1}{\mu_2}D_2\right) + \gamma\left(B_1 + \frac{\mu_1}{\mu_2}B_2\right) + \lambda_1\left(F_1 + \frac{\mu_1}{\mu_2}F_2\right)\right\}, \tag{16}$$

where

$$B_1 = -\alpha \int_0^{x_1}\int_0^{\pi} x^2 \sin^3\theta\, J_{1/2}(\alpha x) J_{3/2}(\alpha x) e^{-\frac{1}{2}\chi^2 \sin^2\theta} e^{-2\lambda_1 x} dx d\theta, \tag{17}$$

$$B_2 = \beta \int_{x_1}^{x_2}\int_0^{\pi} x^2 \sin^3\theta\, [C_2 I_{1/2}(\beta x) + C_3 K_{1/2}(\beta x)][C_2 I_{3/2}(\beta x) - C_3 K3(\beta x)] e^{-\frac{1}{2}\chi^2 \sin^2\theta} e^{-2\lambda_1 x} dx d\theta \tag{18}$$

$$F_1 = \int_0^{\pi} x_1 \sin\theta\, J_{1/2}^2(\alpha x_1) e^{-\frac{1}{2}\chi_1^2 \sin^2\theta} e^{-2\lambda_1 x_1} d\theta, \tag{19}$$

$$F_2 = -\int_0^{\pi} x_1 \sin\theta\, [C_2 I_{1/2}(\beta x_1) + C_3 K_{1/2}(\beta x_1)]^2 e^{-\frac{1}{2}\chi_1^2 \sin^2\theta} e^{-2\lambda_1 x_1} d\theta. \tag{20}$$

## 3. Impurity states in QD in a magnetic field

The Hamiltonian of a QD containing an on-center hydrogenic impurity is given by Eq. (1), with an additional term, which takes into account the Coulomb interaction of the electron

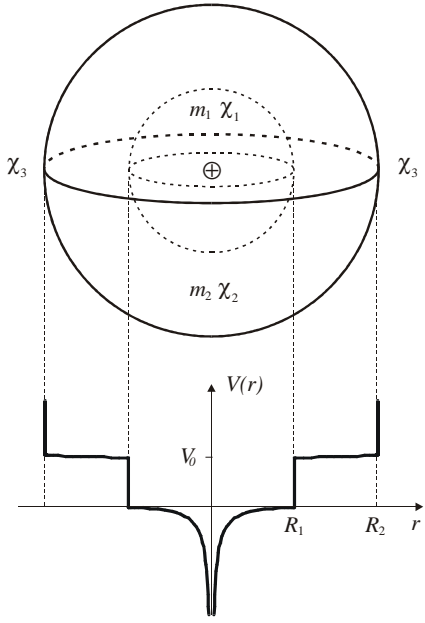

Fig.1 Schematic drawing of the system (for $x = 0.3$, $R_2 = 2R_1$, $\chi_1 = 13.18$ $\chi_2 = 12.244$ $\chi_3 = 10.06$).

with the impurity center in a dielectrically nonuniform system.

Solving the Poisson equation in the sphere, coating and surrounding medium, and using the boundary conditions at the surfaces "sphere-coating" ($r = R_1$) and "coating-surrounding medium" ($r = R_2$), we derive the expression for the potential energy of an electron in the considered system:

$$U(r) = \begin{cases} -\dfrac{e^2}{\chi_1 r} + \dfrac{e^2}{\chi_1 R_1}\left(1 - \dfrac{\chi_1}{\chi_2}\right) + \dfrac{e^2}{\chi_2 R_2}\left(1 - \dfrac{\chi_2}{\chi_3}\right), & r < R_1, \\ -\dfrac{e^2}{\chi_2 r} + \dfrac{e^2}{\chi_2 R_2}\left(1 - \dfrac{\chi_2}{\chi_3}\right), & R_1 \le r \le R_2, \\ -\dfrac{e^2}{\chi_3 r}, & r > R_2, \end{cases} \qquad (21)$$

where $\chi_2$ and $\chi_3$ are the dielectric constants of the coating and the surrounding medium, respectively. When $\chi_1 = \chi_2 = \chi_3$, from Eq. (21) follows the potential energy expression for an electron in the field of a Coulomb center in a dielectricaly homogeneous medium: $U(r) = -e^2/\chi_1 r$. The terms proportional to $(1 - \chi_1/\chi_2)$ and $(1 - \chi_2/\chi_3)$ take into account the difference of DC of the coating layer and surrounding medium. It is necessary to note that $U(r)$ undergoes the greatest change at $\chi_3 = 1$, i.e., when QD is in vacuum.

As in Eqs. (4)-(6), the Hamiltonian of the considered system can be represented as

$$H_i^{imp} = H_{0i} + H_i'^{imp}, \quad (i = 1,2), \qquad (22)$$

where

$$H_{01}'^{imp} = \frac{1}{4}\gamma^2 x^2 \sin^2\theta - \frac{2}{x} - u_1, \qquad u_1 = \frac{2}{x_1}\left(\frac{\chi_1}{\chi_2} - 1\right) + \frac{2}{x_2}\frac{\chi_1}{\chi_2}\left(\frac{\chi_2}{\chi_3} - 1\right), \qquad (23)$$

$$H_{02}'^{imp} = \frac{\mu_1}{\mu_2}\frac{1}{4}\gamma^2 x^2 \sin^2\theta - \frac{2}{x}\frac{\chi_1}{\chi_2} - u_2, \qquad u_2 = \frac{2}{x_2}\frac{\chi_1}{\chi_2}\left(\frac{\chi_2}{\chi_3} - 1\right). \qquad (24)$$

The trial wave function for calculating the ground state energy of the system with the impurity can be chosen as:

$$\psi^{(2)} = \frac{N_2}{\sqrt{x}} e^{-\frac{1}{4}x^2 \sin^2\theta} e^{-\lambda_2 x} \begin{cases} J_{1/2}(\alpha x), & x \le x_1, \\ C_2 I_{1/2}(\beta x) + C_3 K_{1/2}(\beta x), & x_1 \le x \le x_2. \end{cases} \qquad (25)$$

In Eq. (25) $N_2 = (2\pi A_2)^{-1/2}$ is the normalization constant of the wave function, $\lambda_2$ is the variational parameter, $A_2 = D_1^{imp} + D_2^{imp}$, where

$$D_1^{imp} = \int_0^{x_1}\int_0^{\pi} x\sin\theta\, J_{1/2}^2(\alpha x) e^{-\frac{1}{2}\varkappa^2\sin^2\theta} e^{-2\lambda_2 x} dxd\theta, \tag{26}$$

$$D_2^{imp} = \int_{x_1}^{x_2}\int_0^{\pi} x\sin\theta\, [C_2 I_{1/2}(\beta x) + C_3 K_{1/2}(\beta x)]^2 e^{-\frac{1}{2}\varkappa^2\sin^2\theta} e^{-2\lambda_2 x} dxd\theta. \tag{27}$$

Using Eqs. (22), (5), (6) and (23)-(25), after some tedious calculations, we get the following for the impurity ground-state energy:

$$\frac{\varepsilon_1^{imp}}{E_R} = \alpha^2 + \frac{1}{A_2}\left\{(\lambda_2^2+\gamma)\left(D_1^{imp}+\frac{\mu_1}{\mu_2}D_2^{imp}\right)+\gamma\left(B_1^{imp}+\frac{\mu_1}{\mu_2}B_2^{imp}\right)-2\left(G_1^{imp}+\frac{\mu_1}{\mu_2}G_2^{imp}\right)-\right.$$

$$\left. -\left(u_1 D_1^{imp}+u_2 D_2^{imp}\right)+\lambda_2\left(F_1^{imp}+\frac{\mu_1}{\mu_2}F_2^{imp}\right)\right\}, \tag{28}$$

where

$$B_1^{imp} = -\alpha\int_0^{x_1}\int_0^{\pi} x^2\sin^3\theta\, J_{1/2}(\alpha x)J_{3/2}(\alpha x) e^{-\frac{1}{2}\varkappa^2\sin^2\theta} e^{-2\lambda_2 x} dxd\theta \tag{29}$$

$$B_2^{imp} = \beta\int_{x_1}^{x_2}\int_0^{\pi} x^2\sin^3\theta\,[C_2 I_{1/2}(\beta x) + C_3 K_{1/2}(\beta x)][C_2 I_{3/2}(\beta x) - C_3 K3(\beta x)] e^{-\frac{1}{2}\varkappa^2\sin^2\theta} e^{-2\lambda_2 x} dxd\theta \tag{30}$$

$$G_1^{imp} = \int_0^{x_1}\int_0^{\pi} \sin\theta\, J_{1/2}^2(\alpha x) e^{-\frac{1}{2}\varkappa^2\sin^2\theta} e^{-2\lambda_2 x} dxd\theta \tag{31}$$

$$G_2^{imp} = \int_{x_1}^{x_2}\int_0^{\pi} \sin\theta\, [C_2 I_{1/2}(\beta x) + C_3 K_{1/2}(\beta x)]^2 e^{-\frac{1}{2}\varkappa^2\sin^2\theta} e^{-2\lambda_2 x} dxd\theta \tag{32}$$

$$F_1^{imp} = \int_0^{\pi} x_1 \sin\theta\, J_{1/2}^2(\alpha x_1) e^{-\frac{1}{2}\varkappa_1^2\sin^2\theta} e^{-2\lambda_2 x_1} d\theta \tag{33}$$

$$F_2^{imp} = -\int_0^{\pi} x_1 \sin\theta\, [C_2 I_{1/2}(\beta x_1) + C_3 K_{1/2}(\beta x_1)]^2 e^{-\frac{1}{2}\varkappa_1^2\sin^2\theta} e^{-2\lambda_2 x_1} d\theta. \tag{34}$$

The binding energy of the impurity is defined as the difference of the ground state energy of the system without impurity and the ground state energy with impurity: $E_b = \varepsilon_1 - \varepsilon_1^{imp}$.

## 4. Discussion of results

In the numerical calculations carried out for a QD consisting of a *GaAs* sphere coated by a layer of $Ga_{1-x}Al_x As$, embedded in the dielectric medium *AlAs* or in vacuum, the following parameter values have been used [38]: $\mu_1 = 0.067 m_0$, $\mu_2 = (0.067 + 0.083x)m_0$ ($m_0$ is the free electron mass), $V_0 = 1.247 Q_e x$ ($Q_e$ is the conduction band discontinuity fraction), $\chi_1 = 13.18$, $\chi_2 = 13.18 - 3.12x$, $\chi_3 = 10.06$ or 1 for the alloy concentration $x$ within the limits $0 \le x \le 0.4$. In the calculation we neglect the role of the $\Gamma - X$ mixing, which in $GaAs$–$Ga_{1-x}Al_xAs$ systems begins to play a decisive role for values of the radius less than 50Å and $x > 0.5$ [39].

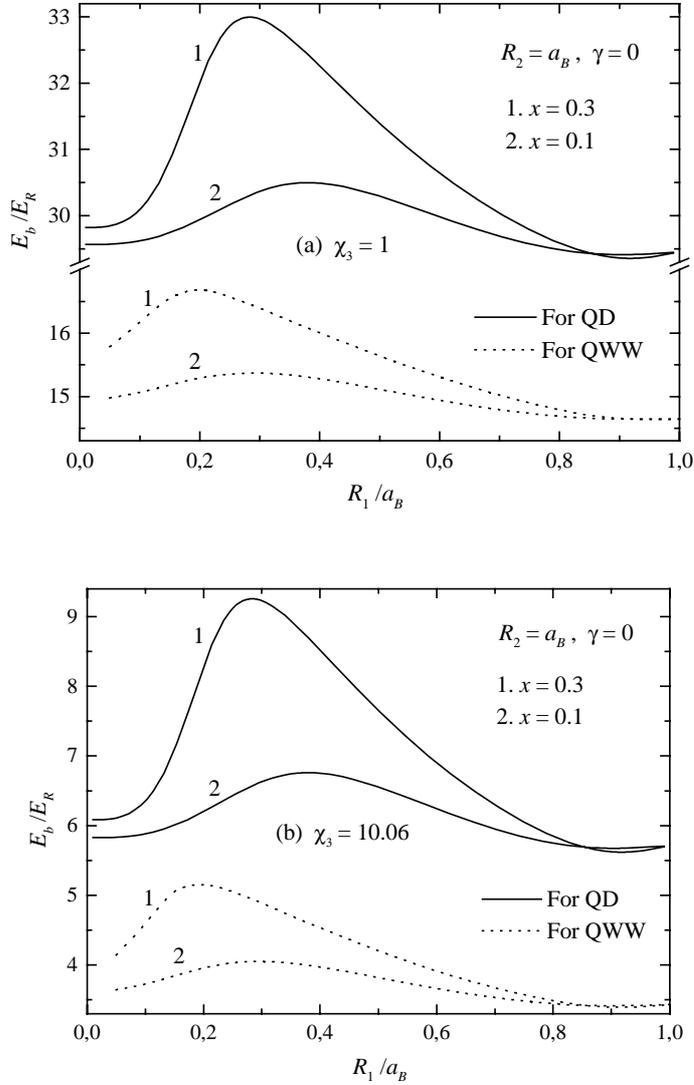

Fig.2. The binding energy dependence on the sphere radius $R_1$ for fixed coating radius $R_2 = a_B$ and for various values of alloy concentration $x$ and DC $\chi_2$, $\chi_3$. (a) OD in vacuum; (b) OD in $AlAs$.

In Fig.2 the dependence of the impurity binding energy on the sphere radius for various values of alloy concentration $x$ and DC $\chi_2$ and $\chi_3$ is presented for a fixed coating radius $R_2 = a_B$, in the absence of magnetic field. From the comparison of curves (1) and (2) it follows that, as the alloy concentration $x$ increases, the maxima of the curves shift to the sphere center. The increase of binding energy is conditioned by the decrease of the electron localization region, as a consequence of increasing potential barrier height at the border of the sphere and the coating, and the strengthening of the system inhomogeneity as a consequence of changes the coating and surrounding environment DC.

From a comparison of the curves in Fig.2 (b), it follows that at $\chi_3 = 10.06$ ($AlAs$) a change of the alloy concentration from 0.1 to 0.3 (the DC of coating decreases about 5%) increases the binding energy by 36.9%. Note that in the case of a wire with the same parameters, the same change of the alloy concentration increases the binding energy by 27% [34].

Because of the small linear dimension of the system ($R_1 \le R_2 = a_B$), the impurity center field is concentrated out of the sphere, essentially in the surrounding environment, so that the DC changes of the environment have a considerable effect on the binding energy. Indeed, if the system is in vacuum (curves of the group (a)), then $\Delta\chi_3 = 9.06$, and the relative change of binding energy at $x=0.3$ equals 3.57, and at $x=0.1$ is about 4.5. The relative change of the binding energy in a wire is 2.24 (at $x=0.3$) and 2.8 (at $x=0.1$) respectively. The observed increase of the relative change of binding energy is conditioned by the electron removal from the sphere center, and the approach to the environment border at the same time. As the sphere radius increases, the effect of DC mismatch on binding energy decreases, and the minima at $R_1 \approx 0.95 a_B$ caused by the effective mass mismatch [40], are smoothed.

The comparison of binding energy values for the wire [33] and QD with the same parameters shows a considerable strengthening of the role of size-quantization in the 3D system.

In Fig.3 the dependence of the impurity center binding energy on the sphere radius is presented for various values of the magnetic field at $x=0.3$, $R_2 = a_B$, $\chi_3 = 1$ (a) and $\chi_3 = 10.06$ (b).

With the increase of $B$, the binding energy increases quickly within the region $R_1 \leq 0.1 \div 0.2 a_B$ and $R_1 \geq 0.7 \div 0.8 a_B$ and slowly in the region $0.2 a_B \leq R_1 \leq 0.7 a_B$. Such behavior of the binding energy is the consequence of the fact that at $0.2 a_B \leq R_1 \leq 0.7 a_B$ the size-quantization prevails over the magnetic one. With the increase of $B$, the depth of the minimum near the border of QD becomes smaller and tends to zero,

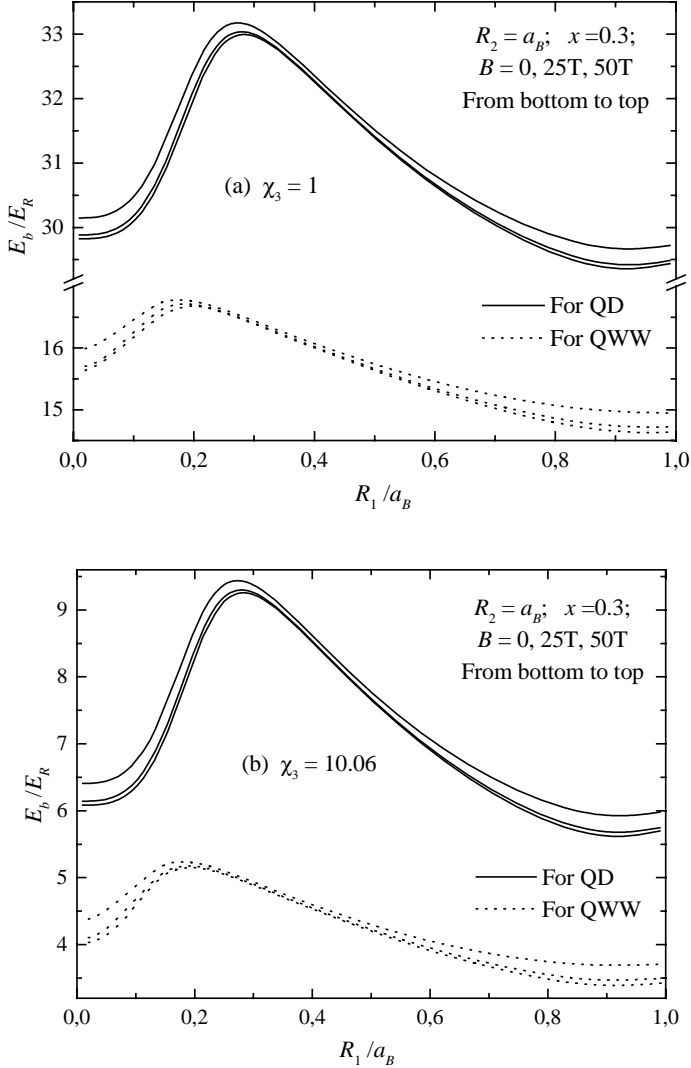

Fig.3. The binding energy dependence on sphere radius $R_1$ for various values of magnetic field at $x=0.3$, $R_2 = a_B$. (a) QD in vacuum; (b) QD in *AlAs*.

because the strong field localizes the electron in the region around the sphere center.

In Fig.4 the dependence of the impurity center binding energy on the sphere radius is presented for various values of magnetic fields at $x=0.3$, $R_2 = 2 a_B$. According to calculations, the location of the maximum (for the given $\gamma$ and $\chi_3$) is shifted to the region of large $R_1$, which is caused by the increase of the linear size of the localization region, conditioned by the removal of the confining barrier. The most important, however, is the effect of dielectric environment, which, although moderates because of the increased coating radius $R_2$, but still considerably affects the binding energy.

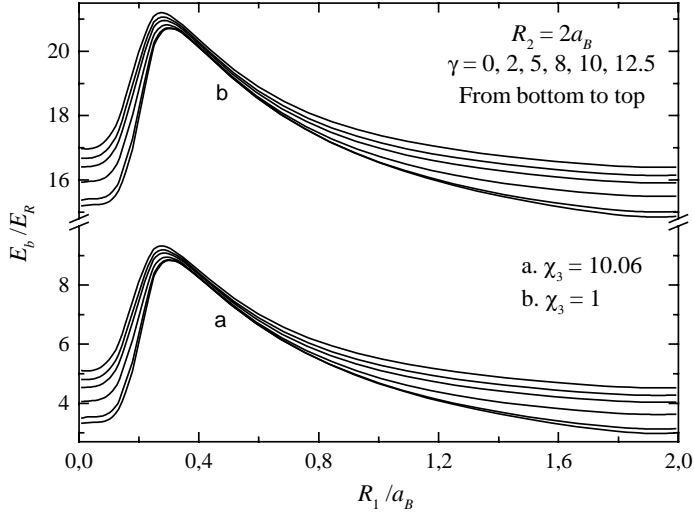

Fig.4. The binding energy dependence on sphere radius $R_1$ for various values of magnetic field at $x = 0.3$, $R_2 = a_B$.

In Fig.5 the dependence of the impurity center binding energy on the magnetic field is presented for a fixed coating radius $R_2 = 2a_B$ and various parameter values of the problem. From a comparison of the curves it follows that, with the increase of sphere radius $R_1$, the dependence of the binding energy on the magnetic field is strengthened. Indeed, according to calculations for "weak" magnetic fields, the dependence of binding energy on $\gamma$ can be presented by the expression $E_b(\gamma) = A + \gamma^2 C$, where the coefficients $A$ and $C$ depend on the values of $R_1$ and $x$. At fixed $x = 0.3$, and for $R_1 = 0.5a_B$, the magnetic field is "weak" when $\gamma << \gamma_0 = (A/C)^{1/2} = 23.6$. With the increase of $R_1$ the parameter $\gamma_0$ decreases, taking the value $\gamma_0 = 9.2$ for $R_1 = a_B$, and $\gamma_0 = 4.7$ for $R_1 = 1.5a_B$. Such behavior of the binding energy is the consequence of the fact that, as was mentioned before, for $0.2a_B \leq R_1 \leq 0.7a_B$ the size-quantization prevails over the magnetic one.

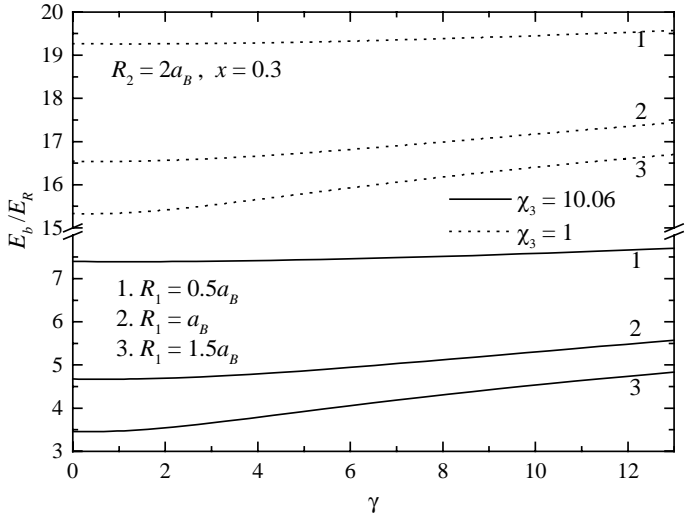

Fig.5. The binding energy dependence on magnetic field for fixed coating radius $R_2 = a_B$, $x = 0.3$ and for various values of sphere radius $R_1$.

## 5. Conclusion

According to the obtained results, the dielectric constant mismatch of the sphere, coating, and surrounding environment appreciably affects the binding energy of the impurity center. This effect increases with the rise of system inhomogeneity, caused both by the increase of the alloy concentration and the decrease of the DC of the coating and environment.

The presence of a magnetic field leads to a rise of binding energy, at that effect increases with both decreasing alloy concentration and decreasing DC of the coating and environment.

The obtained results show that the mentioned effects in the QD's quantitatively exceed the same effects in the QWW's, hence it is necessary to take them into account, especially for small radii of the QD's and relatively large values of alloy concentration.

**Acknowledgments**

This work was supported by the grants INTAS 2001-175 and ISTC A-322.

**Figure captions**

Fig.1 Schematic drawing of the system (for $x = 0.3$, $R_2 = 2R_1$, $\chi_1 = 13.18$ $\chi_2 = 12.244$ $\chi_3 = 10.06$).

Fig.2. The binding energy dependence on the sphere radius $R_1$ for fixed coating radius $R_2 = a_B$ and for various values of alloy concentration $x$ and DC $\chi_2$, $\chi_3$. (a) QD in vacuum; (b) QD in *AlAs*.

Fig.3. The binding energy dependence on sphere radius $R_1$ for various values of magnetic field at $x = 0.3$, $R_2 = a_B$. (a) QD in vacuum; (b) QD in *AlAs*.

Fig.4. The binding energy dependence on sphere radius $R_1$ for various values of magnetic field at $x = 0.3$, $R_2 = a_B$.

Fig.5. The binding energy dependence on magnetic field for fixed coating radius $R_2 = a_B$, $x = 0.3$ and for various values of sphere radius $R_1$.